\documentclass[aps,prd,nofootinbib,twocolumn,showpacs]{revtex4-1}
\usepackage[top=100pt,bottom=0pt,letterpaper]{geometry}
\usepackage{adjustbox}
\usepackage{natbib}
\usepackage[T1]{fontenc}
\usepackage{lmodern}
\usepackage[utf8]{inputenc}
\usepackage{microtype}
\usepackage{comment}
\usepackage{xcolor}
\usepackage[normalem]{ulem}
\usepackage{setspace}
\usepackage{float}
\restylefloat{table}

\usepackage{amsfonts}
\usepackage{appendix}
\usepackage{dcolumn}
\usepackage{lipsum}
\usepackage{graphicx,epsfig}
\usepackage{psfrag}
\usepackage{bm}
\usepackage{epstopdf}
\usepackage{latexsym}
\usepackage{multirow}
\usepackage{amsmath,amssymb}
\usepackage{enumerate}
\usepackage{hyperref}
\usepackage{ulem}
\usepackage[FIGTOPCAP]{subfigure}
\usepackage{empheq,mathtools}
\usepackage{booktabs}
\usepackage{tabularx}
\usepackage{longtable}
\usepackage{tikz}
\usepackage{setspace}
\usepackage{orcidlink}
\def\be {\begin{equation}}
\def\ee {\end{equation}}
\def\ba {\begin{eqnarray}}
\def\ea {\end{eqnarray}}
\def\nn {\nonumber}
\def\bc {\begin{center}}
\def\ec {\end{center}}
\newcommand{\sech}{{\rm sech}}
\newcommand{\bdm}{\begin{displaymath}}
\newcommand{\edm}{\end{displaymath}}

\def\a  {\alpha}
\def\b  {\beta}
\def\g  {\gamma}

\def\D  {\Delta}

\def\l  {\lambda}
\def\m  {\mu}
\def\n  {\nu}
\def\o  {\omega}
\def\O  {\Omega}

\def\r  {\rho}
\def\th {\theta}

\def\s {\sigma}
\def\S {\Sigma}
\def\t  {\tau}

\def\mc {\mathcal}
\def\nn {\nonumber}

\def\ra {\rightarrow}
\def\na {\nabla}

\def\la {\label}
\def\le {\left}
\def\ri {\right}

\def\f {\frac}
\def\sq {\sqrt}
\def\bi {\begin{itemize}}
\def\ei {\end{itemize}}

\def\> {\rangle}
\def\< {\langle}

\def\bc {\begin{center}}
\def\ec {\end{center}}

\begin{document}


\title{Gravitational-wave memory and geodesic-congruence response in generalised Ellis–Bronnikov wormholes}

\author{Soumya Bhattacharya, \orcidlink{0000-0003-2540-7504}}
\affiliation{Department of Physics, Indian Institute of Technology Bombay, Powai, Mumbai - 400076 India}
\email{soumya557@gmail.com}

\author{Suman Ghosh \orcidlink{0000-0002-5270-0135}  }
\affiliation{Department of Physics, Birla Institute of Technology, Mesra, Ranchi-835215, India}
\email{suman.ghosh@bitmesra.ac.in}
%
\begin{abstract}
\noindent We investigate gravitational-wave memory in generalised Ellis–Bronnikov wormhole spacetimes by studying the response of neighbouring timelike geodesics and geodesic congruences to a localized test gravitational-wave pulse. Using proper time as the evolution parameter, we identify pulse-induced displacement and velocity memory after subtracting the background geodesic evolution. We further investigate the response of timelike congruences through the expansion and shear using the Raychaudhuri equations. We find that the memory response depends systematically on the wormhole throat radius, the steepness parameter and the location of the pulse relative to the throat. In particular, a pulse at the throat produces zero net integrated expansion but a finite shear response, while a pulse away from the throat generates an asymmetric congruence response. These results show that wormhole geometry can imprint characteristic signatures on both detector-based and congruence-based memory observables.

\end{abstract}

\maketitle

 
\section{Introduction}

\noindent  Following the detection of gravitational waves (GWs) \cite{LIGOScientific:2016aoc,LIGOScientific:2017vwq},
 and the observations of the shadow of supermassive  compact central  objects, e.g., the M87* and the SgrA*  \cite{EHT:1, EHT:2, EHT:3, EHT:4, EHT:5}, a new era of gravitational physics started.  Both these observations indicate towards the  strong field regime of gravity. They  can be used to test the robustness of general relativity (GR). So far GR not only stands out for its elegance, it keeps on passing experimental tests over the years. There are albeit a few remaining riddles within the theory related to the inability to correctly reconcile singularities occurring either  at the  end
 point of gravitational collapse or at the starting point of our universe. These important shortcomings of GR leave open the scope of working out viable modified theories of gravity. Among the various alternative approaches, one can look at the so called black hole ``{\it mimickers}''
 i.e.  non-black hole compact objects that can arise from quantum gravity-motivated theories, e.g., fuzzballs \cite{Mathur:2005} or compact objects made out of exotic matter fields,  known as exotic compact objects or ECOs \cite{Cardoso:2019,Mark:2017,Cardoso:2016oxy, Murk:2022dkt}. In general, properties of ECOs have not yet been fully explored or understood, nor has it been proven whether or not they exist in nature.  \\
  An important question is whether it is possible to distinguish these so-called black hole mimickers from each other by studying GWs propagating in such background of ECOs. Wormholes \cite{Morris:1988, Bronnikov:2003} are one such class of black hole mimickers that has its origin in general relativity (GR) itself. Although wormholes are speculative objects till date, recent studies of events such as wormhole merger \cite{Krishnendu:2017shb, Cardoso:2016} and their quasinormal modes \cite{Aneesh:2018, DuttaRoy:2019hij}  can illuminate the detection prospects of wormholes in our universe. 

 \noindent One of the most studied traversable wormhole geometry is the four-dimensional Ellis-Bronnikov (EB) spacetime \cite{Ellis:1973, Bronnikov:1973}, which is sustained by a phantom scalar field (a field with a negative kinetic term). 
 The presence of negative energy density makes them difficult to realise in macroscopic quantities \cite{Witten:2019qhl}.
 However, recently, it has been shown that, EB wormhole does partially satisfy energy condition in modified gravity theories such as through embedding in higher dimension \cite{Kar:2015}, \cite{Sharma:2021kqb}.
Further, violation of energy conditions can be avoided at the throat by a simple modification leading to the so-called generalised Ellis-Bronnikov (GEB) model \cite{Kar:1995jz}.
Possible matter source of GEB wormholes has been discussed in \cite{Crispim:2024dgd}. 
Recently, GEB models have constructed within the framework of asymptotically safe gravity \cite{Nilton:2022hrp}, $f(R)$ gravity in \cite{Sokoliuk:2022xcf} and scalar-Einstein-Gauss-Bonnet \cite{Ernazarov:2025fyv}.
Recently, many works have been reported where the EB wormhole spacetime plays a central role, for example the role of phantom scalar field with non-linear electrodynamics in this context is studied in  \cite{crispim2024},  gravitational lensing and deflection angle in \cite{jana:2024}, quasi-normal modes in \cite{Mitra:2023, Khoo_2024}, and in the context of various modified gravity theories \cite{Sharma:2021kqb, Nilton_2023, Sokoliuk_2022}.
However, a comprehensive analysis of the GW memory effect in GEB background has not been reported in the literature.


\noindent The GW memory refers to the lasting change in the relative distance between test particles  when a GW passes through the ambient spacetime \cite{Braginsky:1985, Favata:2010}. 
This effect encompasses both the strong-field, as  well as non-linear aspects of general relativity, which is yet to be observed. 
 It is already proposed to be detected in advanced LIGO and VIRGO \cite{Nichols_2020} and in LISA \cite{LISAConsortiumWaveformWorkingGroup:2023arg}. 
In recent times, there has been a lot of works on various theoretical aspects of memory effect  \cite{Zhang_2017, Horvathy_2017, Zhang_2018, Srijit:2019, Siddhant:2020, Tahura_2021, Nichols_2021, Bhattacharya_2024}. 
In \cite{Bhattacharya:2022, Bhattacharya:2023zel}, it is shown how the signature of the wormhole parameters gets imprinted on the memory signal. 
Recently, it has been emphasized that, in addition to the usual displacement and velocity memory effect, geodesic congruences also experience such an effect known as the $\mc{B}$-memory \cite{OLoughlin:2018ebk}. We discuss about this phenomenon later in more detail.

Our main motivation here is to fill a gap in the existing literature on GEB spacetime by addressing the question: how the memory effect depends on the ambient GEB spacetime. 
To be specific, we investigate displacement-memory, velocity-memory and $\mc B$-memory in the GEB geometry, comparison of throat-crossing and returning geodesics and memory-dependence on the wormhole parameters and
pulse-position. Our analysis is theoretical and qualitative.\\
In Section \ref{sec:review}, we briefly introduce the GEB wormhole model. In Section \ref{sec:geo}, we computed both  displacement and velocity memory effects in a background GEB spacetime. 
We particularly highlight the differences in memory effect when the background evolution is eliminated. 
In Section \ref{sec:B-memory}, we use the Raychaudhuri equation to show that memory effect is visible in the evolution of the geodesic congruences as well. 
We summarise our results with a brief discussion on future avenues in Section \ref{sec:sum}.

\section{GEB geometry and memory set-up} \la{sec:review}
\noindent  It is well known that the EB spacetime can be obtained as an exact solution of the Einstein equations with a phantom scalar field. However, for the GEB wormhole model ($m>2$) additional {\em positive energy} matter is present that prevents the violation of Weak Energy Condition at the throat (and not everywhere). It can be shown that this additional matter satisfies the Average Null Energy Condition \cite{DuttaRoy:2019hij, Kar:1995jz}.
Below we briefly review the geometric set-up of GEB spacetime and operational definition of GW memory.

\subsection{GEB wormhole geometry}

\begin{figure}[!htbp]
\includegraphics[scale=0.4]{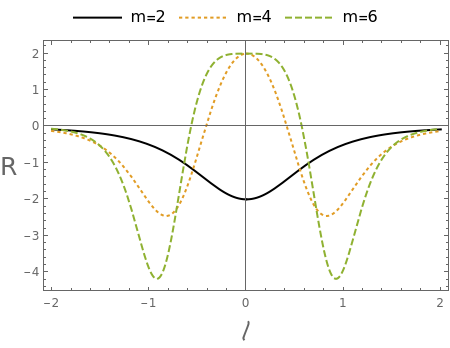}
\caption{\raggedright Ricci scalar vs $l$ for $m=2$ (continuous), $m=4$ (dotted) and $m=6$ (dashed) geometries.}
\label{fig:ricci}
\end{figure}

The GEB wormhole line element \cite{Kar:1995jz} is described in Schwarzschild-like coordinates as, 
\begin{equation}
    ds^2 = - dt^2 + \frac{ dr^2}{1 - \frac{b(r)}{r}  } + r^2 d \Omega_2^2~ ,
\end{equation}
where $d\O_2^2 = d\th^2 + \sin^2\th ^2 d\phi^2$ is the solid angle and the shape function $b(r)$ is given by 
\begin{equation}
    b(r) = r - r^{(3-2m)} (r^m - b_0^m)^{(2-\frac{2}{m})} .
\end{equation}
One gets back the EB geometry, for $m = 2$, which is a static, spherically symmetric, geodesically complete, horizonless space-time (constructed using phantom scalar field). 
Considering the Morris–Thorne conditions \cite{Morris:1988cz} as necessary conditions to construct a Lorentzian wormhole, Kar et al. \cite{Kar:1995jz} introduced the GEB model as a two-parameter ($m$ and $b_0$) family of Lorentzian wormholes.
Here, $b_0$ is the so-called ‘throat radius’ of the wormhole and $m$ is the wormhole `steepness' parameter that can take only even values ($m \ge 2$, to ensure the smoothness of the function $r(l)$).
Let us introduce a tortoise coordinate $l$ such that 
\begin{equation}
    dl = \frac{dr}{\sqrt{\Big(1- \frac{b(r)}{r}}\Big)}.
\end{equation}
In this coordinate the line element looks like 
\begin{equation}
ds^2 = -dt^2 + dl^2 + r(l)^2 \Big(d\theta^2 + \sin^2 \theta d \phi^2  \Big) ,
\label{eqn2}
\end{equation}
where 
\begin{equation}
    r(l) = (b_0^m + l^m)^{\frac{1}{m}}.
\end{equation}
One can see that $l \to \pm \infty$ are the flat asymptotics. 
Corresponding Ricci scalar is given by
\be
R = - \f{2}{r^2} \le(r'^2 + 2 r r'' -1 \ri) .  \la{eq:Ricci}
\ee
Fig \ref{fig:ricci} shows variation of Ricci scalar as a function of $l$, which provides a useful characterization of the curvature profile and of the corresponding matter distribution, as the energy density $\r = R/2$ (see \cite{Sharma:2021kqb} for details). 
It also reflects the maxima of the geodetic potential \cite{Sharma:2022dbx} and the location of the minima of $R$ shows where the negative density matter has accumulated. 
Note that, for $m=2$, about $l \sim 0.63~ b_0$ distance away from the neck (located at $l=0$), the curvature falls below $1/b_0^2$. 
For higher-$m$ geometries the curvature begins to flatten out at even smaller distance from the neck. This implies that, geodesics that pass by at even further distance will feel lesser effect of the wormhole. Therefore, the characteristic size of the wormhole may be effectively identified with the throat radius $b_0$.


\subsection{Test-wave perturbation}

While solving for the memory effect we shall express the line element given in Eq. (\ref{eqn2}) in Bondi-Sachs form as,
\begin{equation}
    ds^2 = -du^2 - 2 du dl + r^2(l)~d\Omega_2^2~,
    \label{BSM}
\end{equation}
where $u=t-l$ is the outgoing light-cone coordinate.
Although the GEB  spacetime approaches Minkowski space as $l \ra \pm \infty$, it is not asymptotically flat in the standard Bondi–Sachs sense. The spacetime possesses two disconnected asymptotic regions, each with its own null infinity. As a consequence, there is no single globally defined retarded time coordinate or Bondi frame covering both sides of the wormhole. This prevents a direct application of the usual BMS-based definition of gravitational-wave memory, which relies on comparing early- and late-time data at the same null infinity \cite{Bhattacharya:2022}. 
One therefore adopts an operational definition of memory based on geodesic and congruence observables, which remains well defined in such multi-asymptotic spacetimes.
We thus use the Bondi–Sachs–like coordinates  purely as a convenient null coordinate system to describe a propagating pulse and do not imply the existence of a standard null infinity. 

The memory effects studied here are defined via permanent changes in geodesic separation and congruence deformation, which remain well defined even in spacetimes with multiple asymptotic regions.
For this purpose, we write down the spacetime metric as a sum of the background wormhole geometry $g_{\mu \nu}$ and a GW perturbation $h_{\mu\nu}$, such that the line element becomes,
\begin{equation}
    ds^2 = (g_{\mu \nu} + h_{\mu \nu}) dx^{\mu} dx^{\nu}
\end{equation}
Using the wormhole geometry presented in Eq. (\ref{BSM}), the resulting geometry becomes,
\begin{equation}
\begin{split}
       ds^2 &= -du^2 - 2 du dl + [r^2(l) + r(l) \,H(u) ]d\theta^2 \\ &+ [r^2(l) - r(l) \,H(u)]\sin^2 \theta d\phi^2~
     \end{split}
     \label{GEBM}
\end{equation}
Here we consider a simplified setup in which we analyze particle motion in this background both in the presence and absence of a pulse represented by $H(u)$ which signifies the brief presence of a wave, noting that the hallmark of GW memory is the lasting effect induced by a transient pulse.
Our analysis will focus on the response of timelike geodesics and geodesic congruences to such a pulse, rather than on constructing a fully self-consistent solution of the perturbed Einstein equations.

Our goal is to isolate geometric signatures of memory effects imprinted by the background wormhole geometry, rather than to model a realistic astrophysical GW source. A self-consistent perturbative treatment will be pursued in future work.
Note that, in flat spacetime analyses of gravitational memory, it is common to use the retarded null coordinate $u$ as an affine parameter. However, in the presence of a curved background and a GW perturbation, $u$ is no longer affine along timelike geodesics. Since gravitational memory is ultimately an observable effect measured by freely falling detectors, we therefore present all results using the proper time $\t$, which carries direct physical significance as discussed in the following subsection.

\subsection{Detector-based interpretation and operational definition of memory }

To connect the geodesic analysis with physically measurable observables, we briefly describe how an idealised gravitational-wave detector would register the memory effects discussed above. We model the detector as a pair (or congruence) of freely falling test masses initially at rest with respect to each other, whose worldlines are timelike geodesics parametrised by their common proper time.
The relative separation vector $\xi^\m$ between neighbouring geodesics obeys the geodesic deviation equation
\be
\f{D\xi^\m}{D\t^2} = R^\m_{\n\a\b} v^\n \xi^\a v^\b  , \la{eq:dev}
\ee
where $v^\a$ is the four-velocity of the detector’s centre-of-mass worldline. The passage of a gravitational-wave pulse produces a transient tidal acceleration, followed by a permanent change in the relative configuration of the detector elements.
In this framework, displacement memory corresponding to a non-vanishing difference can be written as
\be
\na \xi^\m := \xi^\m (\t \ra + \infty ) -  \xi^\m (\t \ra - \infty ) ,
\ee
while the velocity memory corresponds to a residual relative velocity
\be
\na \le(\f{D\xi^\m}{D\t}\ri) \neq 0 .
\ee
Their physical interpretation is obtained from the spatial projection of the separation vector onto the detector's orthonormal frame; the corresponding projected quantities are invariant under coordinate transformations.


\noindent Let us consider two neighbouring geodesics in the wormhole background and then study the evolution of their coordinate separation. 
We confine ourselves in the equatorial ($\theta = \pi/2$) plane and analyse the evolution of geodesic separation along $l ~\& ~\phi$ coordinates. Schematically we can represent them as 
\begin{eqnarray}
    \Delta l \equiv l({\rm Geodesic~II}) - l({\rm Geodesic~I}),~\nonumber\\
\Delta \phi \equiv \phi({\rm Geodesic~II}) - \phi({\rm Geodesic~I}).~\nonumber
\end{eqnarray}
We determine the evolution of the quantities $\Delta l~\& ~\Delta \phi$ both in presence and absence of GW pulse. 
Since memory effect occurs only due to GW, we subtract the background evolution part, if there is any, by defining 
\begin{equation}
    \tilde{\Delta}X(\tau) = \Delta X^{\rm GW}(\tau) - \Delta X(\tau) .
\end{equation}
Here $X$ represents either $l$ or $\phi$ and $\Delta X^{\rm GW}$ represents the separation of two neighbouring geodesics (along $X$) in presence of GW pulse and $\Delta X$ represents separation of two neighbouring geodesics in absence of and GW pulse. 
The new quantity $\tilde{\Delta} X$ extracts the effect of terms (product of $r$ and $H$ and/or their derivatives in equations (\ref{ueq} - \ref{pheq}) ) that couple the wormhole background with the GW pulse.
Thus, we are using the difference between the perturbed and unperturbed geodesic separations to isolate the contribution induced by the GW pulse. The asymptotic change of this pulse-induced separation is then used as a diagnostic of displacement memory.
If the geodesics do not have constant separation, after the passage of the GW pulse, one can associate a velocity memory with these geodesics as well. 
Therefore the coordinate representation of the velocity memory effect can be expressed as
\be
\tilde{\Delta} v_X = \lim_{\t \ra +\infty} \f{d}{d\t} \tilde{\Delta} X - \lim_{\t \ra -\infty} \f{d}{d\t} \tilde{\Delta} X . \la{eq:vel-mem-def}
\ee
Below we proceed to the analysis of GW memory.

\section{Memory effect through geodesic analysis} \la{sec:geo}

\noindent In this section, we present the analysis of the memory effect vis-a-vis the geodesic deviation between neighbouring geodesics due to a propagating GW pulse, with the geodesic separation quantifying the amount of displacement memory effect. 
We have solved the three geodesic equations, presented in Eq. (\ref{ueq}) - Eq. (\ref{pheq}), numerically using the standard software package {\it MATHEMATICA} and have presented the solutions graphically. 
\noindent The geodesic equations for the line element given in Eq. (\ref{GEBM}) for $u$, $l$, $\theta$ and $\phi$ coordinates as
\begin{eqnarray}
   \ddot{u} + \frac{r'}{2}\Big(2 r + H\Big) \dot{\theta}^2 +  \frac{r'}{2}\Big(2 r - H\Big) \sin^2 \theta \dot{\phi}^2  = 0. ~~ \label{ueq} 
\end{eqnarray}
\ba
 \Ddot{l} - \frac{1}{2}\Big(2 r r' +  r' H - r H'\Big) \Dot{\theta}^2 && \nonumber \\  -\frac{1}{2}\Big( 2 r r' -  r' H + r H' \Big) \sin^2 \theta \Dot{\phi}^2 && = 0. \hspace{0.3in}
 \label{leq}
\ea
\ba
  \Ddot{\theta} &+&  \f{H'}{H + r} \dot{u}\dot{\th}  + \frac{(2r  + H)r' }{(r  + H)r}\Dot{l}  \Dot{\theta}  \nn \\
 &-&  \frac{ r - H }{ r + H} \sin \theta \cos \theta \Dot{\phi}^2 = 0. \hspace{0.3in} \label{theq}
\ea

\ba
  \Ddot{\phi} &-& \frac{H'}{r - H} ~\Dot{u} \Dot{\phi}  + \frac{(2r - H)r'}{(r - H)r} ~\Dot{l}\Dot{\phi} \nn \\
  &+& 2 \cot \theta ~\Dot{\theta} \Dot{\phi} = 0. \label{pheq}
\ea

\noindent Here an ‘overdot’ denotes derivative
with respect to the proper time $\tau$ associated with the geodesics, while ‘prime’ denotes derivative with respect to the argument of the respective function. For example, $H'(u) \equiv (dH/du)$ and $r'(l) \equiv (dr/dl)$. 
Equations \ref{ueq}-\ref{pheq} suggest that, for the radial geodesics considered here, the chosen GW perturbation does not enter the geodesic equations, and hence these particular geodesics do not exhibit a pulse-induced memory signal.
However, this is not the case for geodesic congruences which we shall exploit in the next section.

\noindent We consider the time-like geodesics that satisfy following geodesic constraint
\begin{eqnarray}
     -\dot{u}^2 - 2~\dot{u} \dot{l} + [r^2(l) + r(l)~H(u) ]~\dot{\theta}^2 && \nonumber \\ + [r^2(l) - r(l)~ H(u)]\sin^2 \theta ~\dot{\phi}^2  = &&-1 .\hspace{0.3in}
     \label{tbl}
\end{eqnarray}
In flat backgrounds, it is standard practice to present memory effect graphically with evolving $u$ as it is an affine parameter and $u=0$ presents the location of the pulse as well. 
Strictly speaking, in presence of the pulse, $u$ is not an affine parameter and  we present all results using the proper time $\t$, which has direct operational meaning as discussed earlier.
We choose the GW test pulse profile as
\begin{equation}
    H(u)=A~ \sech^2 (u/u_0),  \label{wave}
\end{equation}
which is localised at $u=0$. Here, $u_0$ indicates the pulse width. For all computational purposes, we have set $u_0=1$. Note that, positive definiteness of the metric component $g_{\phi\phi}$ implies $r>H$ which further implies $b_0>A$ (i.e. the pulse is only a small perturbation to the background wormhole geometry). For numerical evaluation, we choose $A=1/2$ and $b_0>1/2$.
In the following, we analyse two scenarios where geodesics meet the pulse at different locations-- at and near the throat.

\subsection{Memory effect on throat-crossing geodesics when the pulse is near the throat}

\noindent 
Here we study the case where two `throat-crossing' geodesics meet the pulse near the throat.
We choose our initial conditions for the two geodesics as 
\begin{eqnarray}
    {\rm Geodesic~I:} ~~u(0)=0, ~l(0)=0,~\phi(0)=2.0, \nonumber\\ \dot{u}(0)=1,~\dot{l}(0)=1.~~~~~~~~~~~~~~
\end{eqnarray}
\begin{eqnarray}
    {\rm Geodesic~II:} ~~u(0)=0, ~l(0)=0.1,~\phi(0)=2.0, \nonumber\\ \dot{u}(0)=1,~\dot{l}(0)=1.~~~~~~~~
\end{eqnarray}

\begin{figure}[!htbp]
\includegraphics[scale=0.4]{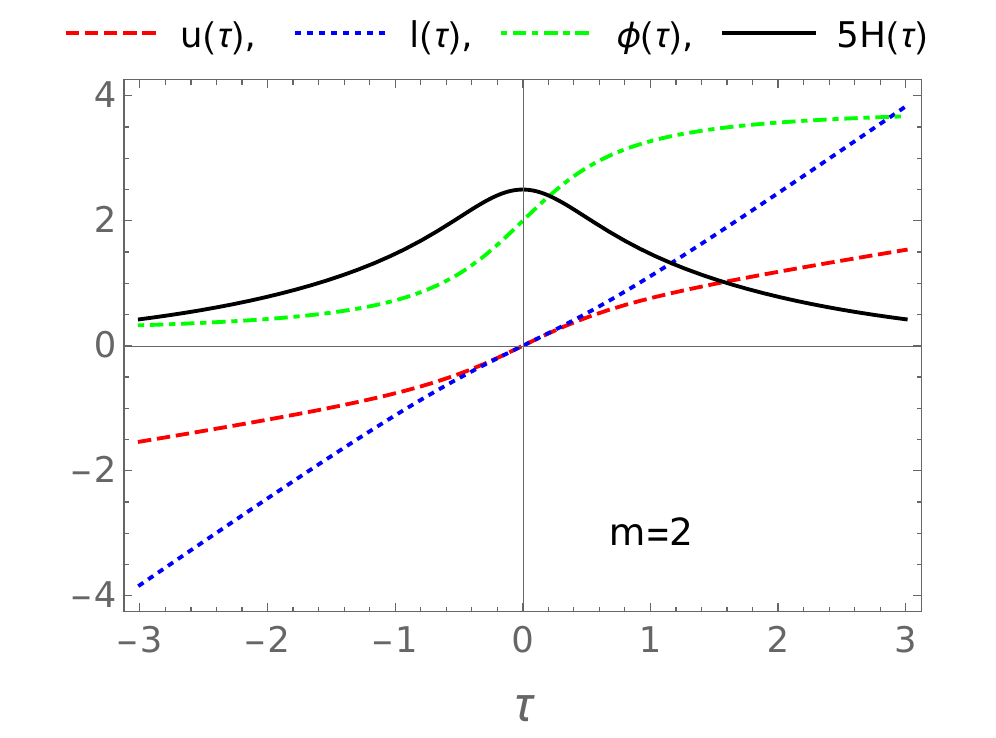}
\caption{\raggedright Evolution of geodesics and the (scaled) pulse profile with respect to the proper time $\t$ for $m=2$ (EB) model.}
\la{fig:geo-m2-l0}
\end{figure}
Figure \ref{fig:geo-m2-l0} shows that $u(\t)$ varies monotonically with $\t$ thus justifying our choice of representing the memory effect with respect to the proper time $\t$ instead of $u$ (which would be appropriate in a flat background).
Note that these geodesics traverse through the wormhole throat (from negative to positive $l$-coordinates).
Further, the pulse peaks at $\t =0$ and the geodesic meets the pulse at the throat $l=0$.


\begin{figure}[!htbp]
\includegraphics[scale=0.4]{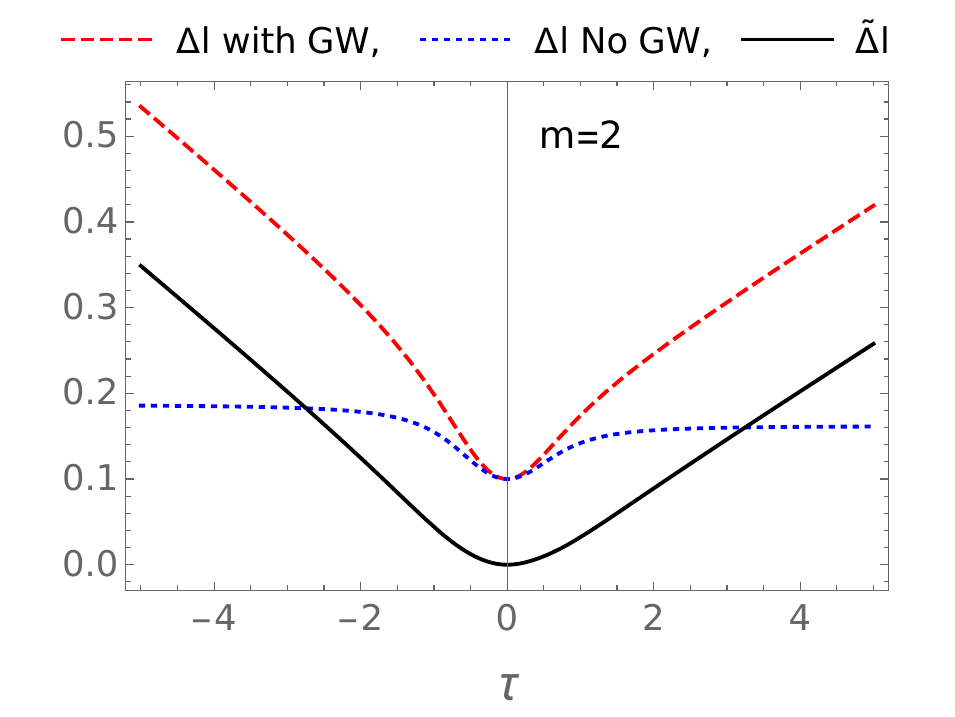}
\caption{\raggedright Evolution of the $\Delta l$  in presence (red/ dashed curve), in absence (blue/dotted curve) of GW  and  $\tilde{\Delta}l$ (black/continuous curve) for $m=2$.}
\la{fig:m2-l}
\end{figure}
Figure \ref{fig:m2-l} show the evolution of the separation of geodesics in presence (red/dashed line), in absence (blue/dotted line) of the pulse and also $\tilde \D l$ that captures the effect of the pulse modulo the background evolution.

\begin{figure}[!htbp]
\includegraphics[scale=0.55]{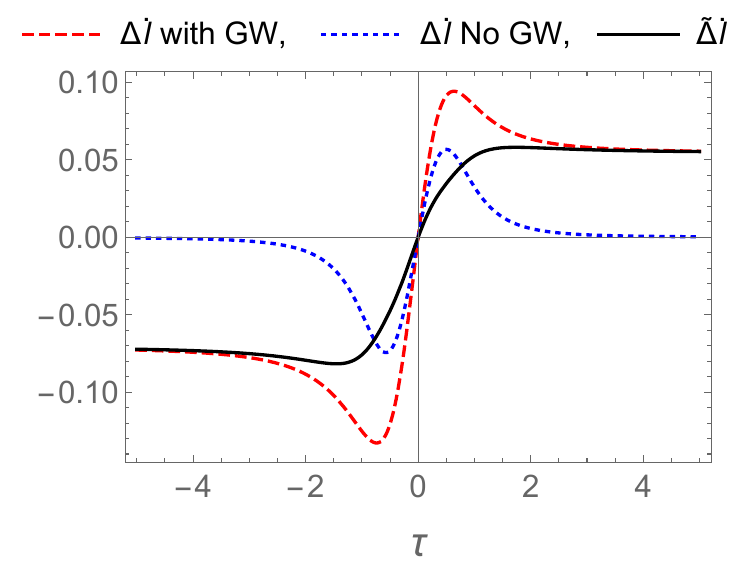}
\caption{\raggedright Evolution of the $\D \dot{l}'$  in presence (red/ dashed curve), in absence (blue/dotted curve) of GW  and  $\tilde{\D}\dot{l}$ (black/continuous curve) for $m=2$.}
\la{fig:m2-l-vel}
\end{figure}
Fig. \ref{fig:m2-l-vel} show the evolution of the difference of geodesic velocities along $l$-direction, in presence (red/dashed line), in absence (blue/dotted line) of the pulse and also $\tilde \D \dot{l}$.
Interestingly, Fig. \ref{fig:m2-l-vel} clearly shows two geodesics which have asymptotically different velocity difference.

\begin{figure}[!htbp]
\includegraphics[scale=0.4]{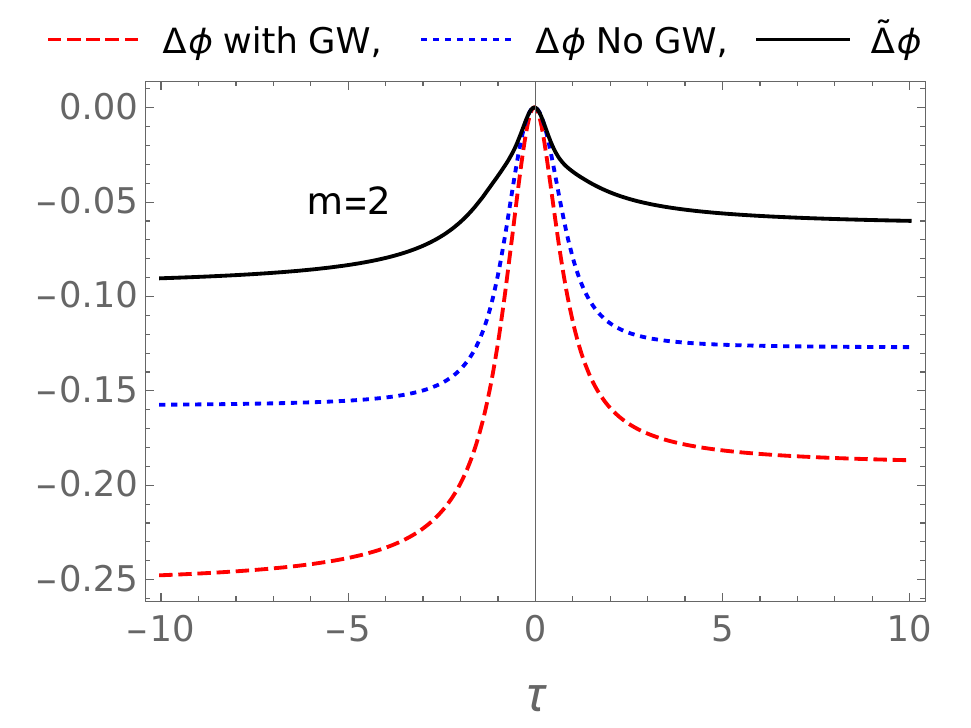}
\caption{\raggedright Evolution of the $\Delta \phi$  in presence (red/ dashed curve), in absence (blue/dotted curve) of GW  and  $\tilde{\Delta}\phi$ (black/continuous curve) for $m=2$.}
\la{fig:m2-phi}
\end{figure}
Figures \ref{fig:m2-phi} and \ref{fig:m2-phi-vel} show the evolution of difference in $\phi$-coordinate and the corresponding velocities respectively. There is indication of displacement memory but no apparent signature of velocity memory effect between asymptotic regions along angular coordinate.

\begin{figure}[!htbp]
\includegraphics[scale=0.5]{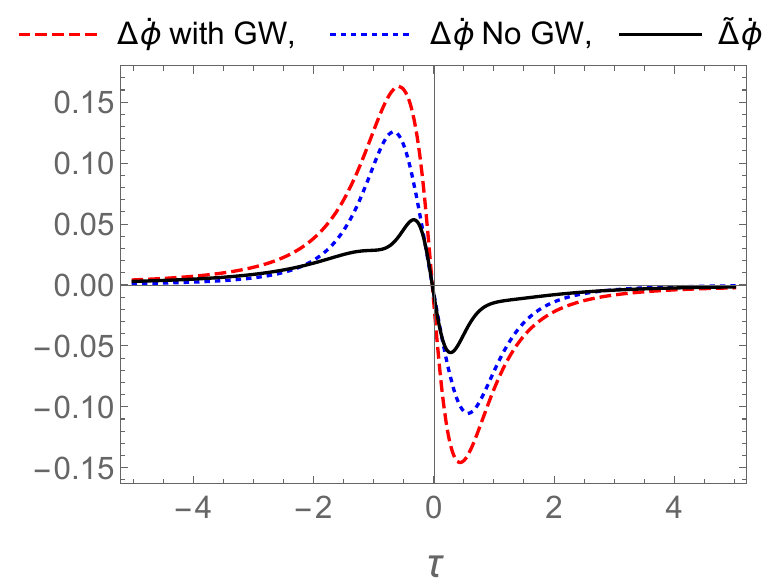}
\caption{\raggedright Evolution of the $\D \dot{\phi}$  in presence (red/ dashed curve), in absence (blue/dotted curve) of GW  and  $\tilde{\D}\dot{\phi}$ (black/continuous curve) for $m=2$.}
\la{fig:m2-phi-vel}
\end{figure}
Below we discuss the scenario where the pulse meets the geodesics away from the throat.

\subsection{Memory effect on returning geodesics and the pulse is away from the throat}

In the following, we analyse the scenario where two geodesics meet the pulse away from the throat (but not too far as that would be similar to the flat background scenario). This analysis is similar to the one performed in \cite{Bhattacharya:2022}.
We choose our initial conditions for the two geodesics as 
\begin{eqnarray}
    {\rm Geodesic~I:} ~~u(0)=5.0, ~l(0)=2.0,~\phi(0)=2.0, \nonumber\\ \dot{u}(0)=1,~\dot{l}(0)=1.~~~~~~~
\end{eqnarray}
\begin{eqnarray}
    {\rm Geodesic~II:} ~~u(0)=5.0, ~l(0)=2.5,~\phi(0)=2.0, \nonumber\\ \dot{u}(0)=1,~\dot{l}(0)=1.~~~~~~
\end{eqnarray}
Boundary conditions are chosen such that the pulse is at $\t \sim 2$. For both the cases, the initial values of $\dot{\phi}$ are determined through eq \ref{tbl}. 
Note that these geodesics are by construction {\em returning} \cite{Sharma:2022dbx} type i.e. coming from infinite distance from the throat. They get to a point of closest approach and then return to infinity.


\begin{figure}[!htbp]
\includegraphics[scale=0.66]{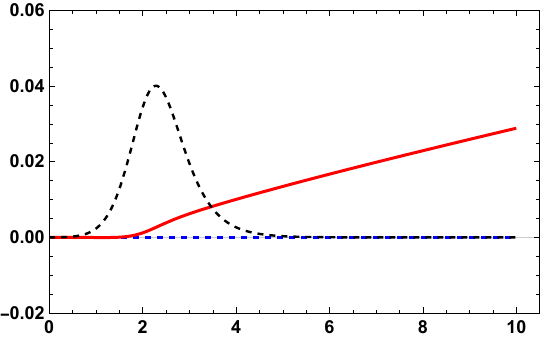}
\put(-180,90){$\bf {\tilde{\Delta} l}$}
\put(-10,-10){$\bf {\tau}$}
\caption{\raggedright Evolution of $\tilde{\Delta} l$ in presence (red/ continuous curve) and absence (blue/ dashed curve) of GW (bell-shaped curve) for $m=2$.}
\label{memory_l_d}
\end{figure}
\begin{figure}[!htbp]
\includegraphics[scale=0.66]{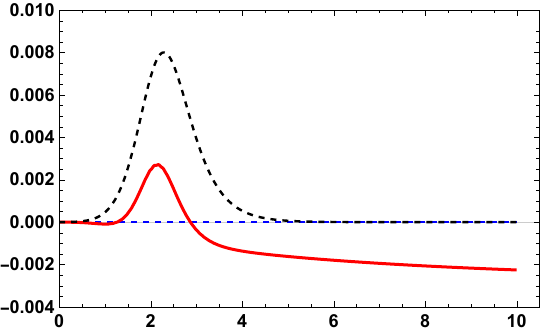}
\put(-190,90){$\bf {\tilde{\Delta } \phi}$}
\put(-10,-10){$\bf {\tau}$}
\caption{\raggedright Evolution of the $\tilde{\Delta}\phi$  in presence (red/ continuous curve) and absence (blue/ dashed curve) of GW (bell-shaped curve) for $m=2$ (EB) model.}
\label{memory_p_d}
\end{figure}

The figures \ref{memory_l_d} and \ref{memory_p_d} represent the evolution of $\tilde{\Delta}l$ and $\tilde{\Delta}\phi$ (for $m=2$) respectively. In both the figures, the GW pulse is represented by the black dotted curve. 
Here one can clearly see that after the passage of the pulse there is a permanent change in the evolution of both $\tilde{\Delta}l$ and $\tilde{\Delta}\phi$. This is the displacement memory effect. 
Different slopes of these curves in the asymptotic region are indicative of presence of velocity memory as well.


\noindent After showing the presence memory effect qualitatively, in the following, we discuss how variation of the wormhole parameters modify the memory effect.

\subsection{Parameter dependence: varying the $A/b_0$ ratio and $m$}

\begin{figure}[!htbp]
\includegraphics[scale=0.66]{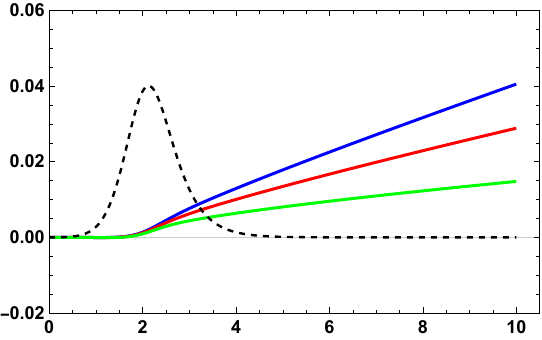}
\put(-180,90){$\bf {\tilde \Delta}l $}
\put(-10,-10){$\bf {\tau}$}
\caption{\raggedright Evolution of $\tilde{\Delta}l$ for $b_0=1.0$ (red/middle curve), $b_0=0.8$ (blue/top curve) and $b_0=1.2$ (green/bottom curve) in $m=2$ model.}
\label{memory_comp_l_b}
\end{figure}

\begin{figure}[!htbp]
\includegraphics[scale=0.66]{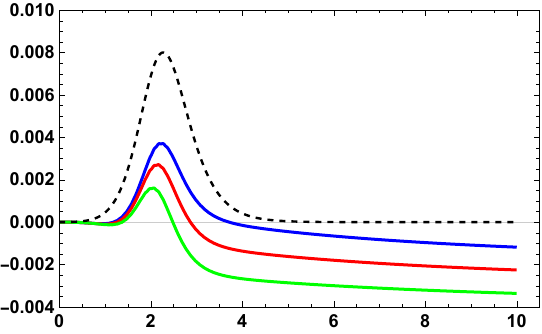}
\put(-190,90){$\bf {\tilde{\Delta} \phi}$}
\put(-10,-10){$\bf {\tau}$}
\caption{\raggedright Evolution of $\tilde{\Delta}\phi$  for $b_0=1.0$ (red/middle curve), $b_0=0.8$ (blue/top curve) and $b_0=1.2$ (green/bottom) in $m=2$ model.}
\label{memory_comp_p_b}
\end{figure}

\begin{figure}[!htbp]
\includegraphics[scale=0.66]{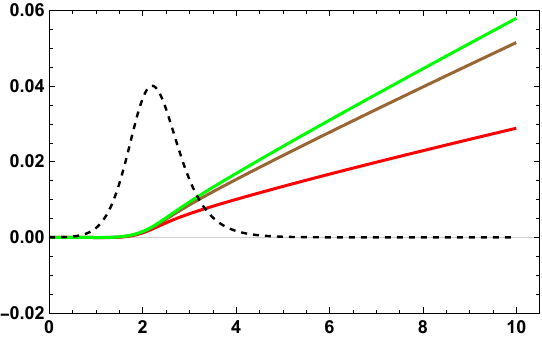}
\put(-180,90){$\bf {\tilde{\Delta} l}$}
\put(-10,-10){$\bf {\tau}$}
\caption{\raggedright Evolution of the $\tilde{\Delta} l$ for $m=2$ (red/bottom), $m=4$ (brown/middle) and $m=6$ (green/top) model keeping $A/b_0=1/2$ fixed.}
\label{memory_comp_l_1}
\end{figure}

\begin{figure}[!htbp]
\includegraphics[scale=0.66]{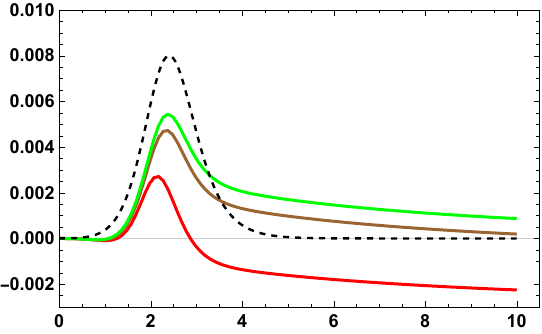}
\put(-190,90){$\bf {\tilde{\Delta} \phi}$}
\put(-10,-10){$\bf {\tau}$}
\caption{\raggedright Evolution of the $\tilde{\Delta}\phi$  for $m=2$ (red/bottom), $m=4$ (brown/middle) and $m=6$ (green/top) model keeping $A/b_0=1/2$ fixed.}
\label{memory_comp_p_1}
\end{figure}

Here we analyse the dependence of memory effect on varying wormhole parameters. We consider only the returning geodesics as presented in Section III-B. Parameters dependence we presented below are generic, i.e. throat-crossing geodesics also show similar dependency.
Figures \ref{memory_comp_l_b}-\ref{memory_comp_p_b} show that decreasing the throat radius $b_0$ (and $A=\f{1}{2}$), while keeping the steepness parameter $m$ fixed, leads to a scenario similar to increasing $m$ while keeping $b_0$ fixed as shown in figures \ref{memory_comp_l_1}-\ref{memory_comp_p_1}. 
Note that for both set of changes scalar curvature of the spacetime geometry decreases at any particular $l$. 


In case of Fig. \ref{memory_comp_l_b}, as we make $b_0<1$ (blue) keeping $A$ fixed, then the magnitude of the negative-curvature contribution from the GEB background becomes larger relative to the curvature induced by the pulse. 
Opposite effect can be seen with increasing $b_0$.

On the other hand, in Fig. \ref{memory_comp_p_b}, apart from the above effect, what we also see is a signature of both the wormhole and pulse geometry being spherically symmetric (thus independent of $\phi$).
In the following we keep $A/b_0 =1/2$ and vary the wormhole parameter $m$.
In Fig. \ref{memory_comp_l_1}-\ref{memory_comp_p_1}, something slightly different happens which can be understood as follows. 
As fig. \ref{fig:ricci} suggests for $m>2$, throat has a positive Ricci scalar and the (negative) minima appears closer to $l=b_0$. 
As the pulse meets the geodesics at $l=2$, in higher $m$-geometries, geodesics move further away.

\begin{table*}[t]
\centering
  \renewcommand{\arraystretch}{1.4} 
  \setlength{\tabcolsep}{6pt}     
  \begin{tabular}{|c|c|c|c|c|c|c|}
    \hline
Figures   & Geometry & Pulse at & Shift in $\tilde{\D}l$ & Shift in $\tilde{\D}\dot{l}$ & Shift in $\tilde{\D}\phi$ & Shift in $\tilde{\D}\dot{\phi}$ \\ \hline \hline
 \ref{fig:m2-l}, \ref{fig:m2-l-vel}, \ref{fig:m2-phi}, \ref{fig:m2-phi-vel} & $m=2$ &  Throat &  Finite   &  Finite & Finite & Zero \\ \hline
 \ref{memory_l_d}, \ref{memory_p_d} & $m=2$ &  Away &  Finite   &  Finite & Finite & Small/non-zero  \\ \hline
 \ref{memory_comp_l_b}, \ref{memory_comp_p_b} & Decreasing $A/b_0$  &  Away &  Decreases   &  Decreases & Increases & Zero  \\ \hline
 \ref{memory_comp_l_1}, \ref{memory_comp_p_1} & Increasing $m$ &  Away &  Increases   & Decreases & Increases & Zero  \\ \hline  
  \end{tabular}
  \caption{Summary of numerical analysis presented in Section \ref{sec:geo}} \la{tab:geo}
\end{table*}


\noindent Results that can be qualitatively inferred, within numerical resolution, from the plots in this section is summarised in Table \ref{tab:geo} (where by ``Shift'', we essentially mean change in asymptotic values of $ \tilde{\D}X $). 
After discussing the effects of the wormhole parameters on the memory effect, in the following section, we analyse the GW memory of geodesic congruences in GEB background.


\section{Geodesic congruence response: Diagnostic $\mc{B}$-memory } \la{sec:B-memory}

Solving geodesic equations and geodesic deviation equation leads to possible displacement memory effect in general.  However, geodesic equation can reveal velocity memory as well.
The memory effect involving geodesic congruence is closer to the velocity memory but not quite the same.
A covariant approach of looking at the behavior of geodesic congruences has been proposed recently by O’Loughlin and Demirchian \cite{OLoughlin:2018ebk}, wherein the term $\mc{B}$-memory ($\mc{B}$ denotes the covariant derivative of the velocity field) is introduced in the context of impulsive gravitational waves-- 
A permanent change in the kinematics of the geodesic congruences, characterized covariantly through the $B_{\a\b}$-tensor (defined below).
The fundamental $\mc{B}$-memory object in the literature is the change in the congruence tensor $B_{\a\b}$ whose evolution is described by the Raychaudhuri equations.

In \cite{Chak:2020}, Raychaudhuri equation is used to study the memory effect on a congruence of geodesics, in exact plane wave spacetimes expressed in the Brinkmann coordinates.
They found that for a growth (or decay) of shear causes focusing of an initially parallel congruence, after the departure of the pulse. 
Remarkably, a correlation between the ``focusing time'' and the amplitude and phase of the pulse (or its derivatives) is found. Such features distinctly suggest presence of the so-called $\mc{B}$-memory.
Thus, for extended objects, $\mc{B}$-memory provides a complementary covariant characterization of the GW memory effect.
In general curved background, comparing geodesics could be difficult as the individual affine parameters are independent of each other.
Remarkably, a congruence of geodesics, by construction, is parameterised by a single affine parameter.

\subsection{Congruence formalism: the Raychaudhuri equations}

One may derive the ESR variables directly from the following definitions \cite{Wald:1984, Poisson:2009pwt, Kar:2006ms, Kar:2008zz, Samuel:2020fim, Shaikh:2013iea} of the expansion scalar $\theta$, the shear tensor $\Sigma_{\a\b}$ and the rotation tensor  $\Omega_{\a\b}$, by first solving for the geodesic vector field $v^\a$, 
\begin{eqnarray}
\theta &=& \nabla_\a v^\a  \label{eq:thetadef},\\
\Sigma_{\a\b} &=& \frac{1}{2} \left (\nabla_\b v_\a +\nabla_\a v_\b\right ) -\frac{1}{n-1} h_{\a\b}\theta \label{eq:sigmadef}, ~~\\
\Omega_{\a\b}&=&\frac{1}{2} \left ( \nabla_\b v_\a - \nabla_\a v_\b \right ).\label{eq:omegadef}
\end{eqnarray}
Here, $n$ is the dimension of spacetime and $h_{\a\b} = g_{\a\b} \pm v_\a v_\b$ is the projection tensor (the plus sign is for timelike curves whereas the minus one is for spacelike ones) and $v_\a v^\a =\mp 1$. 

\noindent One may also solve the following equation \cite{Wald:1984}
\begin{equation}
v^\m\nabla_\m \mc{B}_{\a\b} = - \mc{B}_{\a\m}\mc{B}^\m_{\,\b} - R_{\a\m\b\n} v^\m v^\n , \label{eq:gen}
\end{equation}
where the $\mc{B}$-tensor is given by, 
\be
\mc{B}_{\a\b}= \nabla_\b v_\a = \f{1}{n-1}\th h_{\a\b} + \s_{\a\b} + \o_{\a\b}, \la{eq:B-ESR}
\ee
along with the geodesic equations.
Then it is easy to determine the so-called ESR variables from $\mc B_{\a\b}$ using the definitions in Eqs. (\ref{eq:thetadef} - \ref{eq:omegadef}).
One can separate Eq. (\ref{eq:gen}), into the following three equations for trace, antisymmetric and symmetric traceless parts of $\mc B_{\a\b}$ as,
\be
\f{d\th}{d\l} + \frac{1}{n-1} \th^2 + \s^2 - \o^2 = - R_{\a\b} v^\a v^\b  ,\la{eq:theta}
\ee
\ba
\f{d\s_{\a\b}}{d\l} = -\f{2}{n-1} \th \s_{\a\b} - \s_{\a\g} \s^\g_{~\b} + \o_{\a\g} \o^\g_{~\b} \nn \\ 
- \frac{(\s^2 - \o^2)}{n-1} h_{\a\b} + C_{\m\b\a\n} v^\m v^\n + \f{\tilde{R}_{\a\b}}{n-1}  ,\la{eq:sigma}
\ea
\be
\f{d\o_{\a\b}}{d\l} = -\f{2}{n-1} \th~ \o_{\a\b} - \s^\g_{~\b} \o_{\a\g} +  \s^\g_{~\a} \o_{\b\g},\la{eq:omega}
\ee
where, $\s^2 = \s_{\a\b}\s^{\a\b}$, $\o^2 = \o_{\a\b}\o^{\a\b}$ and $C_{\m\b\a\n}$ is the Weyl tensor and $\tilde{R}_{\a\b} = h_{\a\m}h_{\n\b} R^{\m\n} - \f{1}{n-1} h_{\a\b}h_{\m\n} R^{\m\n}$.
Note that, these are not equations but, identities. In some scenarios \cite{Frankel:1979, Carter:1996wr, Zafiris:1997he}, however, we ﬁnd that the identities become equations once we use the Einstein equations or any other geometric property (e.g. Einstein space, or vacuum, etc.) as an input.
To see the evolution from a local observer's viewpoint \cite{Ghosh:2010gq}, one can solve the geodesic deviation equation too and express the tensorial quantities in the frame basis. The metric tensor in the coordinate basis and frame basis are related as
\begin{equation}
g^{\a\b} = e^\a_{\,\,a}\,e^\b_{\,\,b}\, \eta^{ab},
\end{equation}
where the vierbein field, $e^\a_a$, has two indices, `$\a$' labels the general spacetime coordinate (the coordinate basis) and `$a$' labels the local Lorentz spacetime or local laboratory coordinates (the frame basis). However, we do not present the evolution with respect to the local observer and only look at the evolution of ESR variables by numerically solving Eq. (\ref{eq:gen}) as it is the asymptotic features that are more relevant in the context of the memory effect.

\subsection{${\mc B}$-memory observable}

For an extended detector or a continuous distribution of test particles, similar information is encoded in the expansion ($\th$, discussed later) and shear of the associated timelike geodesic congruence. In particular, a non-vanishing integrated expansion, $\int d \t ~ \th(\t)$, signals a permanent deformation of the detector volume and thus the integrated expansion provides a scalar diagnostic/measure of the $\mc{B}$-memory associated with the congruence, even when the instantaneous expansion returns to zero at late times.
For a timelike congruence
\be
 \th(\t) = \f{1}{\mc A} \f{d {\mc A}}{d\t}
\ee
where ${\mc A}$ is the appropriate cross-sectional volume/area element. Therefore
\be
\int_{\t_i}^{\t_f} d \t ~ \th(\t) = \log \f{\mc A_f}{\mc A_i}
\ee
This `area under the curve' of the $\th$-profile represents the net logarithmic change in the cross-sectional area of a geodesic bundle \cite{Hochberg:1997wp, Kar:2006ms} as it travels through a wormhole as such.
Thus one may consider the following quantity as a measure of $\mc{B}$-memory
\be
\mc{M}_{\th} = \int \le(\th_{GW} - \th_0\ri) d\t , \la{eq:E-mem}
\ee
where $\th_{GW}(\t)$ and $\th_0(\t)$ are expansion scalar in presence and in absence of the GW pulse respectively. 
Similarly, shear-memory can be defined as
\be
\mc{M}_{\S^2} = \int \le(\S^2_{GW} - \S^2_0\ri) d\t . \la{eq:S-mem}
\ee
These quantities are used here as scalar diagnostics of the pulse-induced change in the congruence kinematics; they should not be regarded as substitutes for the full tensorial $\mc{B}$-memory.
Further, these memory effects correspond to genuine, detector-based observables rather than coordinate artifacts, and remain well defined in spacetimes lacking a global null infinity.

In the following, by solving the Raychaudhuri equation, we reveal qualitative features of the memory effect in the evolution of a geodesic congruence in GEB spacetimes (for various choices of the wormhole parameters), when a gravitational wave pulse passes through it. 
We present the evolution of expansion and shear with respect to $l$.
Recall that both $l(\t)$ and $u(\t)$ evolve monotonically with $\t$ (as shown in Fig. \ref{fig:geo-m2-l0}) and thus can be used interchangeably. 


\subsection{${\mc B}$-memory: throat-crossing congruence}

\begin{figure}[!htbp]
\includegraphics[scale=0.35]{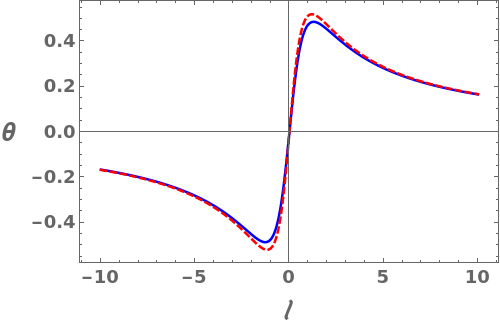}
\includegraphics[scale=0.35]{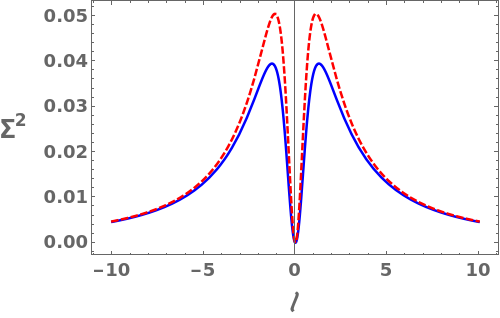}
\caption{\raggedright Expansion scalar $\th$ and the shear amplitude $\S^2$ of congruence of timelike geodesics for $m=2$ model with GW pulse (dashed lines) and without GW pulse (continuous lines).  }
\label{fig:theta-sigma-m2}
\end{figure}

We present, in Fig. \ref{fig:theta-sigma-m2}, the evolution of the expansion scalar and the shear amplitude for congruence of timelike geodesics that are crossing the EB wormhole ($m=2$) throat from positive to negative $l$. 
Here we assume a non-rotating congruence for simplicity. Boundary conditions are set such that $\mc{B}_{\a\b}(l=0)=0$ i.e., disappearing expansion and shear when the congruence crosses the wormhole throat (i.e. $l=0$ at $\t = 0$). 
 We note that the boundary conditions chosen here—vanishing expansion and shear at the throat—represent a special class of congruences designed to highlight the imprint of the gravitational pulse. 
 We do not claim that the resulting ${\mc B}$-memory is universal for all timelike congruences. Rather, we demonstrate that a suitably chosen congruence acquires a pulse-dependent change in its kinematic invariants.
 While alternative boundary conditions may alter the quantitative details of the evolution, the present analysis demonstrates the qualitative existence of $\mc{B}$-memory in GEB wormhole spacetimes. A systematic exploration of generic congruences will be addressed in future work.\\
For the `central' geodesic, boundary conditions are chosen as-- $\dot{u}(0)=\sq{2} -1, \dot{l}(0) = 1, \dot{\th}(0)=\dot{\phi}(0)=0$ and $u(0)=0, l(0) = 0, \th(0)=\pi/2 , \phi(0)=0$.
Thus, the central geodesic, which is a radial geodesic, does not see any effect of the gravitational pulse as such, which is clear from the geodesic equations.
Note that it is the curvature term in Eq. (\ref{eq:gen}) that determines the evolution of the congruence.
In the figures, the continuous and dotted lines represent congruences in absence and presence of the gravitational pulse, respectively. 
The initial conditions for the pulse are chosen so that the pulse reaches the throat at the same time that the congruence arrives. 
For all numerical estimations, we have chosen $b_0 = 1, A=1/2$ like in the previous section. Note that the results in the absence of the pulse matches with the results reported earlier \cite{DuttaRoy:2019hij,Sharma:2022dbx} (where the non-null coordinates were used), thus assuring accuracy of our numerical evaluation.

Fig. \ref{fig:theta-sigma-m2} shows that on a background EB spacetime (i.e. $m=2$), the pulse results in increased  expansion rate and shear deformation of the congruence for $l>0$ and increased compression rate for $l<0$. 
This can be understood as follows. 
Note that, in the $l \ra \infty$ limit, $\th$ vanishes for both with and without GW pulse. 
At the throat of the wormhole, expansion shifts from negative (converging geodesics) to positive (diverging geodesics), acting as a defocusing region.
Presence of the GW pulse enhances this shift.
Fig. \ref{fig:theta-sigma-m2} suggests that the area under the curve or net expansion of the cross section of geodesic congruences vanishes both in presence and absence of GW pulse.  
Further $\mc{M}_{\th}$ also vanishes identically.
This is what we mentioned earlier about radial geodesics in Sec III.
Interestingly, the shear $\S^2$ is an even function of $l$. Thus the area under the curve gives us a non-zero net shape deformation of the cross-section. 
There is a visible difference, in the area under the curve of the shear-profile, in presence and in absence of the GW pulse-- which suggests memory effect on the shape of the congruence i.e a non-zero $\mc{M}_{\S^2}$.
Thus, even when the pulse produces no net integrated expansion of the congruence, it can leave a finite shear-induced deformation, demonstrating that the memory response need not be reducible to a net volume change.

\begin{figure}[!htbp]
\includegraphics[scale=0.36]{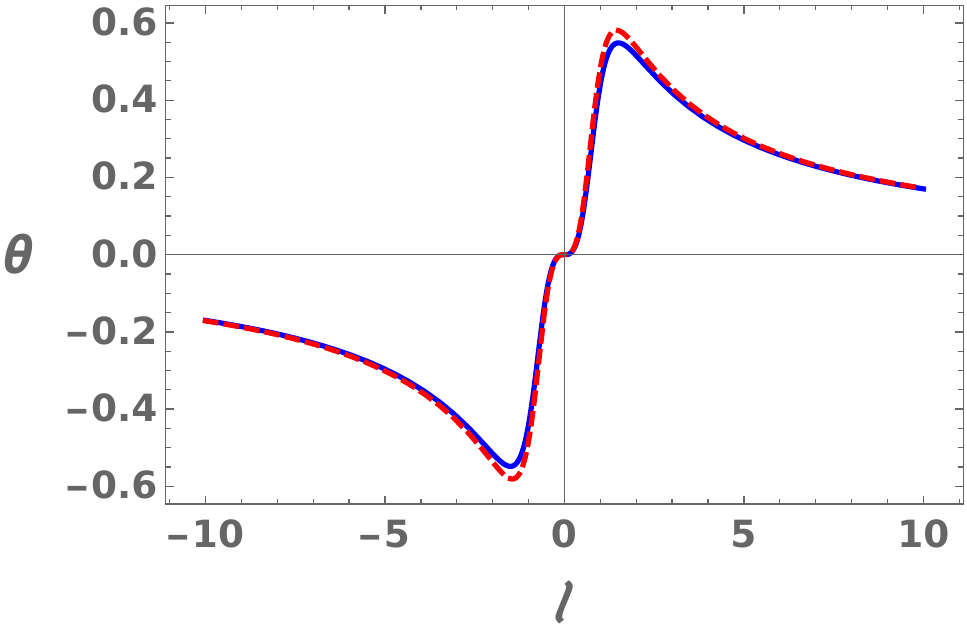}
\includegraphics[scale=0.35]{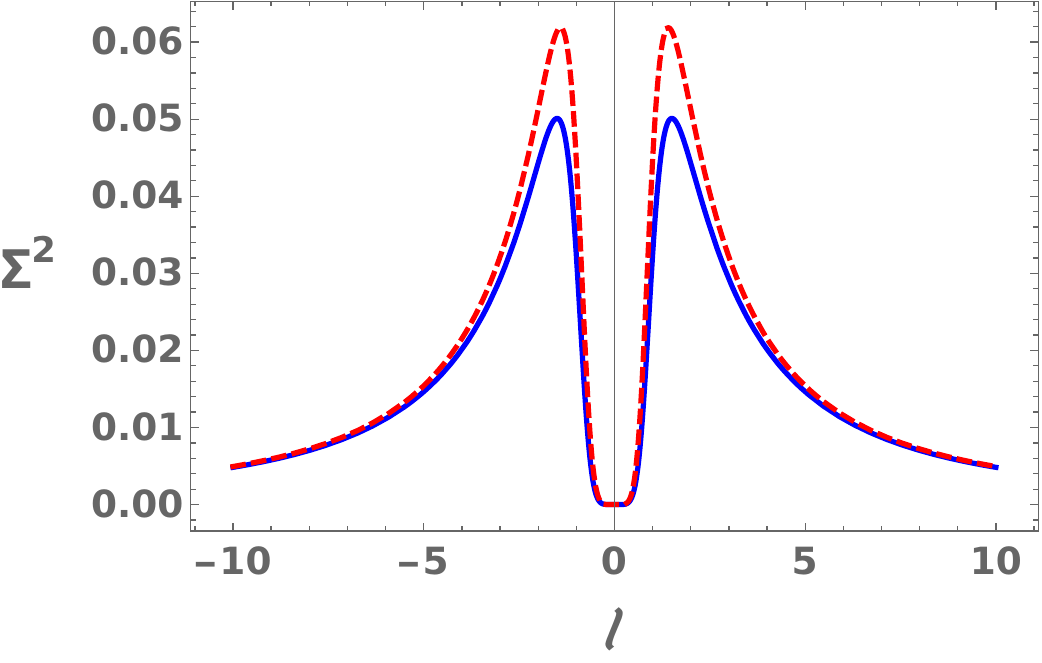}
\caption{\raggedright Expansion scalar $\th$ and the shear amplitude $\S^2$ of congruence of timelike geodesics for $m=4$ model with GW pulse (dashed lines) and without GW pulse (continuous lines).}
\label{fig:theta-sigma-m4}
\end{figure}

Fig. \ref{fig:theta-sigma-m4} shows similar features for $m=4$ (and for higher $m$-geometries). 
However, the amount of shift in $\th$ across the throat decreases slightly with increasing $m$.
This is because for higher $m$, the region associated with the exotic matter supporting the wormhole becomes increasingly localized away from the throat (see Fig. \ref{fig:ricci} and \cite{Sharma:2021kqb}), thus weakening the defocusing effect.
Geometrically speaking, geodetic potential has a {\em flatter} maxima near the throat for higher $m$.
As a clear feature of higher-$m$ geometry (irrespective of pulse-effect), we see that the selected congruence exhibits a weaker pulse-induced deformation for larger $m$, indicating reduced sensitivity of this memory observable to the pulse.
We refer to this as a greater kinematic stability of the throat-crossing congruence.
We observe that the location, where the effect of the pulse on $\th$ is maximum, approaches $l = b_0$ as we increase the value of $m$, however, variation of $\th$ itself peaks at $l=0$. 

It is worth mentioning that in plane wave spacetime, a GW pulse induces momentary gravitational attraction between two test bodies, moving parallel to each other, causing the corresponding geodesics to converge in finite time. 
Such convergence would be visible both in displacement memory and $\mc{B}$-memory \cite{Chak:2020}. 
In case of wormholes, expansion rate $\th$ does not diverge to negative infinity, rather goes to zero asymptotically \cite{DuttaRoy:2019hij, Sharma:2022dbx}. 
This is expected for any wormhole spacetime. Even the presence of the pulse, that carries a positive energy density, fails to overcome the effect due to its weaker nature compared to the exotic sources of the GEB wormhole. 
In a `realistic' scenario, one would expect such pulse to generate near the throat from weak perturbation of the wormhole.

\noindent We note that, exact one-to-one comparison between results of this section with Sec. III is difficult as $\th$ essentially represents a four dimensional divergence of the velocity four-vector. Such tracing loses some information about memory effect along particular direction as presented in Sec III. 
Further the geodesics analysed in the previous section are finitely apart, whereas a congruence is made up of infinitesimally nearby geodesics.
Below we analyse the scenario where the congruence meets the pulse away from the throat and the effect of varying  $A/b_0$ ratio on $\mc{B}$-memory for throat-crossing geodesic congruences.

\subsection{When pulse is away from the throat}

\begin{figure}[!htbp]
\includegraphics[scale=0.36]{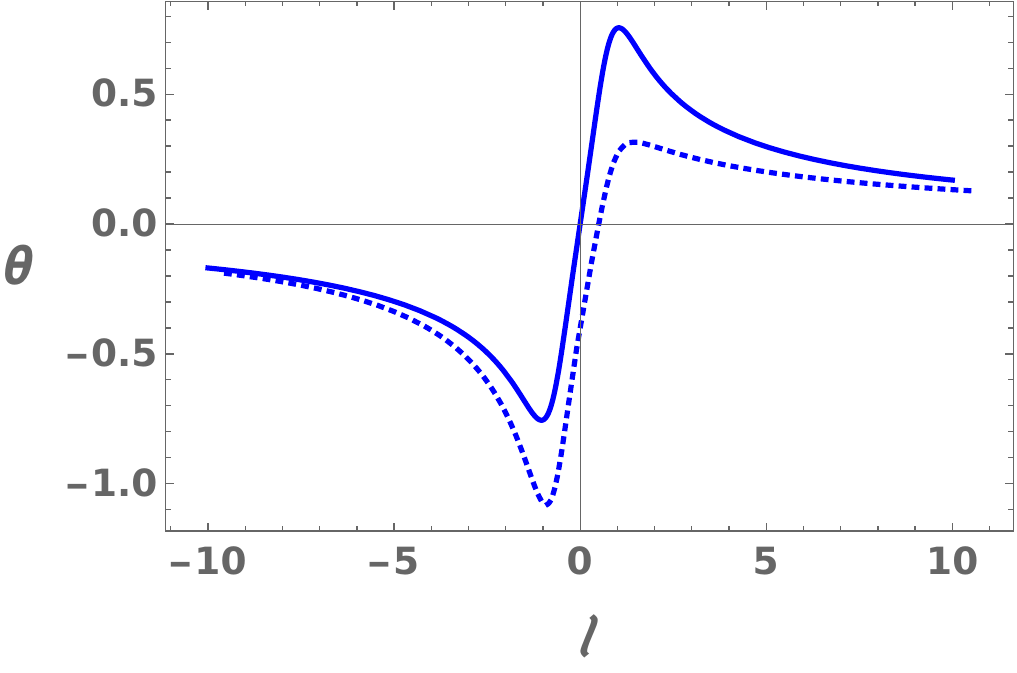}
\includegraphics[scale=0.36]{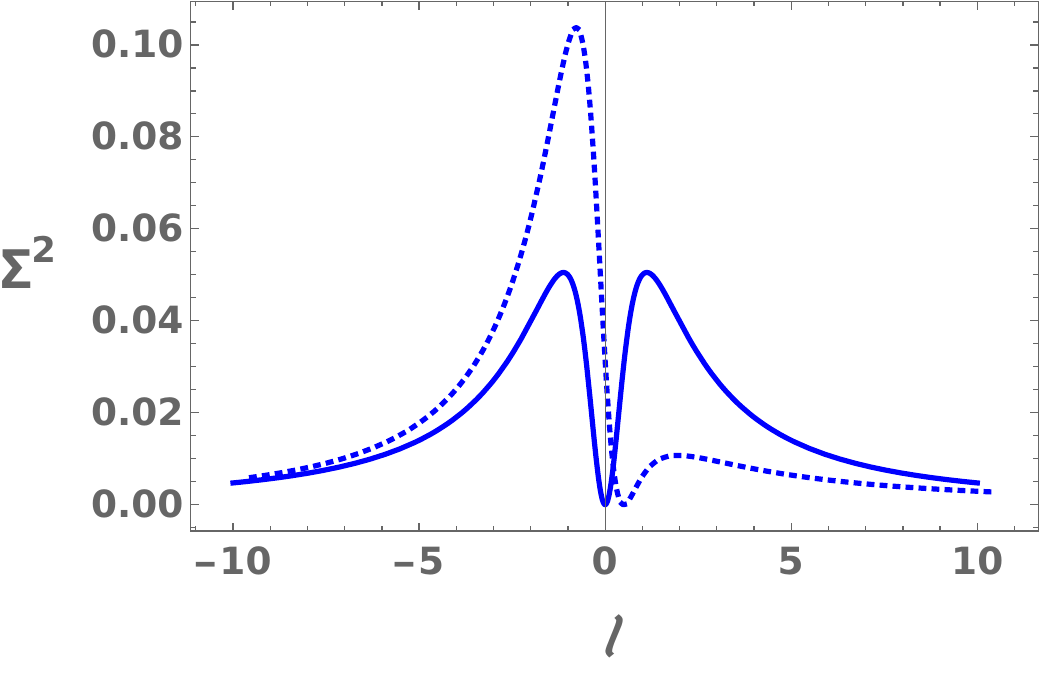}
\caption{\raggedright Expansion and shear profiles (for $m=2$) when the congruence meets the pulse at $l=0$ (continuous curve) and at $l=1/2$ (dotted curve).}
\label{fig:vary-meet-m2}
\end{figure}

In Fig. \ref{fig:vary-meet-m2}, we show the evolution of the expansion scalar and shear (vs. $l$) when the geodesics meet the pulse at $l=1/2$ to compare it with the case where the congruence and the pulse meet at the throat ($l=0$). We have set the boundary conditions such that the congruence of radial geodesics meets the radially moving pulse at $l=1/2$ (other boundary conditions are same as in Fig. \ref{fig:theta-sigma-m2}).  
Fig. \ref{fig:vary-meet-m2} shows that the $\th$-profile (dotted curve) has a net non-zero (negative) expansion when the pulse is met away from the throat compared to the case where the congruence meets the pulse at the throat and has zero total expansion (continuous curve). 
Similarly, if we choose the `meeting point' to be $l=-1/2$, we will have equal but positive net expansion. Further, the presence of the pulse at $l=1/2$ decreases the rate of expansion or defocusing. 
Thus, presence of a pulse away from the throat essentially breaks the symmetry across the throat ($l=0$) and the congruence effectively sees an asymmetric wormhole.
The shear profile also shows a considerable difference in the two cases.
Thus the wormhole background converts the location of the pulse relative to the throat into an observable asymmetry in the congruence response and the pulse position dependence is one of the most interesting results we report.



\subsection{Parameter dependence}

\begin{figure}[!htbp]
\includegraphics[scale=0.34]{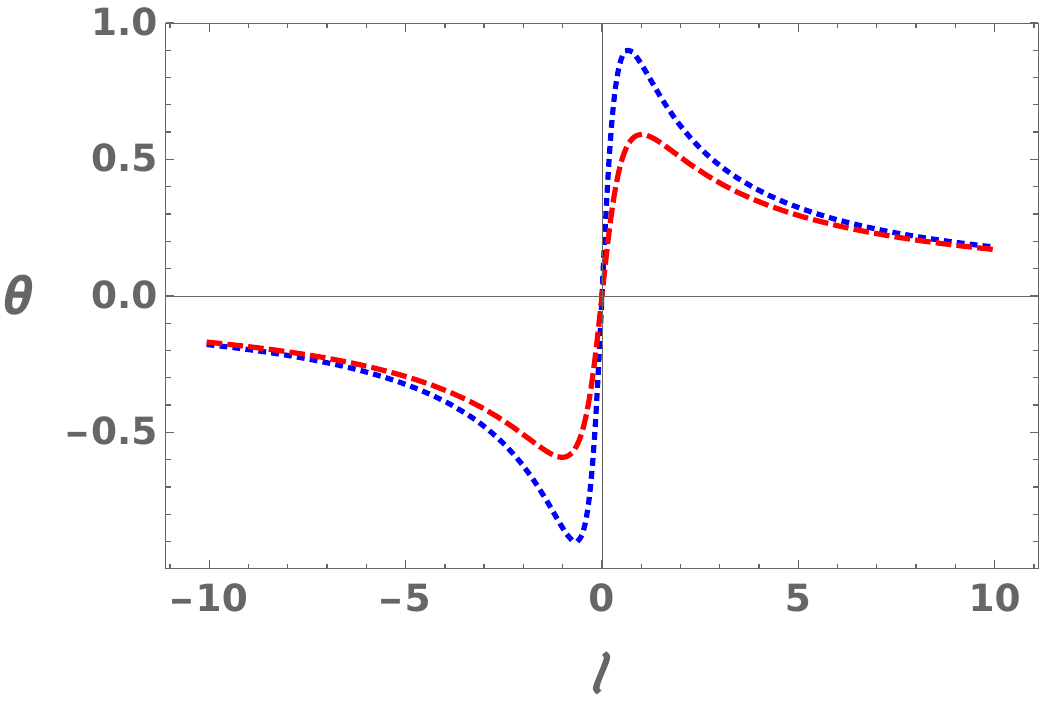}
\includegraphics[scale=0.35]{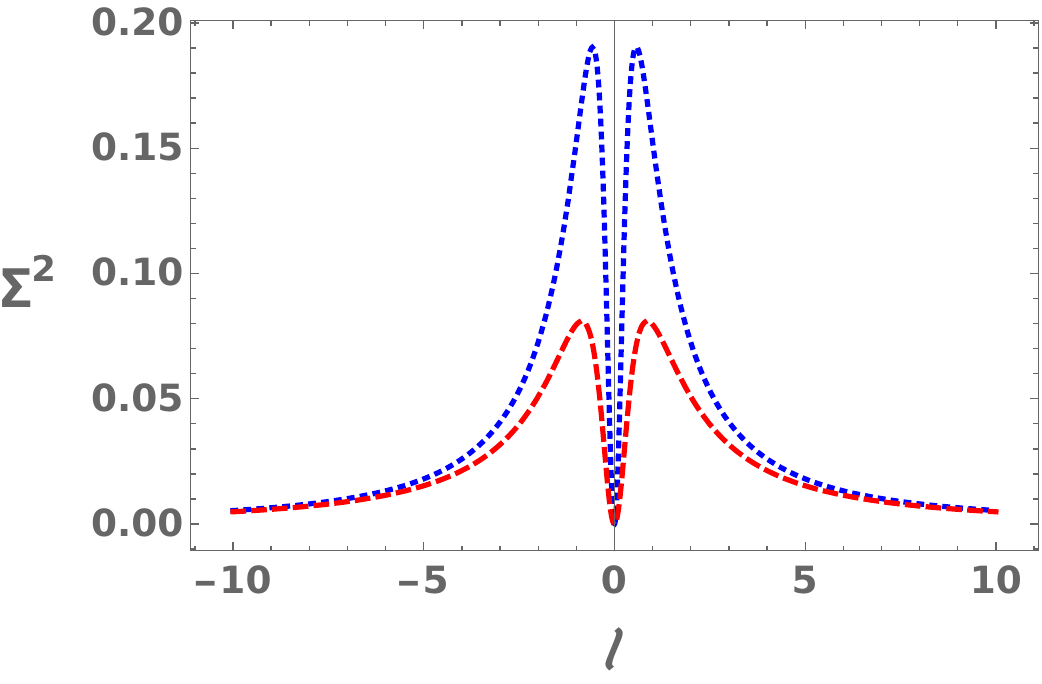}
\caption{\raggedright Expansion and shear profiles for throat radii $b_0=1$ (dashed curves) and $b_0=2/3$ (dotted curves) in presence of GW pulse ($A=1/2$).}
\label{fig:vary-b}
\end{figure}

Fig. \ref{fig:vary-b} shows the evolution of expansion and shear for a congruence of `throat-crossing' geodesics in EB spacetime ($m=2$) in presence of a pulse (at the throat) with varying throat radius (and a fixed pulse profile $A=1/2$). These patterns can be understood as follows. 
As we lower the length scale of the wormhole compared to the length scale of the pulse (which is fixed), the pulse becomes increasingly dominant in determining the evolution of the geodesic congruence thus imparting a larger effect. However, the total expansion or the area under the curve remains zero as expected from previous discussions. 
On the other hand, we see significant increase in shear as the throat radius gets closer to the pulse amplitude.
We have tabulated the qualitative trends from the numerical results of Section \ref{sec:B-memory} in Table \ref{tab:B-memory}.

\begin{table*}[t]
  \centering
  \renewcommand{\arraystretch}{1.4} 
  \setlength{\tabcolsep}{5pt}     
  \begin{tabular}{|c|c|c|c|c|c|c|}
    \hline
Figure   & Geometry & Pulse at & $\th_{GW} - \th_0$ &  $\mc{M}_{\th}$ & $\S^2_{GW} - \S^2_0$ & $\mc{M}_{\S^2}$ \\ \hline \hline
 \ref{fig:theta-sigma-m2}, \ref{fig:theta-sigma-m4} & $m=2,4$ &  Throat &  Skew-symmetric   &  Zero & symmetric & Finite \\ \hline
 \ref{fig:vary-meet-m2} & $m=2$  &  $l=\pm 1/2$ &  Asymmetric  & $\pm$ Finite & Highly asymmetric & Finite  \\ \hline
\ref{fig:vary-b}  & Increasing $A/b_0$ &  Throat &  Increases & Zero & Increases & Increases  \\ \hline  
  \end{tabular}
  \caption{Summary of numerical analysis presented in Section \ref{sec:B-memory}} \la{tab:B-memory}
\end{table*} 

\section{Discussions} \la{sec:sum}

\noindent In this article, we have explored GW memory effect in the generalised Ellis-Bronnikov wormhole spacetime. 
We, at first, have briefly reviewed the geometry of the  GEB spacetime.  
Then we have explored aspects of GW memory by analyzing  neighboring geodesics and the geodesic congruences in the wormhole background in presence of a localized GW pulse. 
We found that the GEB wormhole and the pulse geometry modify both detector-based displacement/velocity memory and the kinematic response of timelike geodesic congruences, with the magnitude and character of the response depending on the wormhole throat scale, steepness parameter, and location of the gravitational-wave pulse.
Our results should be interpreted within the test-wave approximation adopted here; the pulse is not obtained by solving the linearized Einstein equations around the GEB background. A self-consistent perturbative treatment will be explored elsewhere.
Let us summarize, the significant findings of the paper.

\begin{itemize}

\item We have shown, explicitly how the geodesic separation evolves in the presence of the GW pulse, modulo the background effect. 
This provides evidence for a pulse-induced displacement memory in the operational geodesic-separation sense adopted here.
The different late-time asymptotic slopes provide evidence for a nonzero residual relative velocity, consistent with velocity memory.
Also we have shown through various plots how the memory effect depends on the wormhole {\it hairs}, $b_0$ and $m$, and provided physical explanation of the numerical results in Sec III. 
 
\item We analysed further the aspects of memory effect on the geodesic congruences-- the so-called $\mc{B}$-memory-- using the Raychaudhuri equations. 
We have analysed geodesic congruence that crosses the wormhole throat towards positive $l$-direction. 
The $\mc{B}$-memory does not exactly mimic displacement or velocity memory as expected. Evolution of expansion and shear shows clear difference in total deformation of the congruence in presence of the pulse compared to the scenario without the pulse.
We showed that the expansion profile depends crucially on where the congruence meets the pulse.
The GEB background can also distinguish volume-type deformation from shape-type deformation.

\item Since the expansion scalar represents a rate rather than a cumulative deformation, we argue that, the area under the expansion curve captures the total expansion experienced by the congruence and serves as a useful estimator of $\mc{B}$-memory.
A quantitative measure of {\em expansion}-memory is thus provided by the integrated expansion scalar (modulo background effect) between asymptotic regions. 
We found, in particular, that the presence of a GW pulse away from the throat essentially reflects in the memory effect.
The pulse breaks the reflection symmetry of the dynamical response, so that the congruence effectively experiences an asymmetric geometry despite the underlying GEB background remaining reflection symmetric.

\item  $\mc{B}$-memory also depend crucially on the {\it hairs} or the wormhole parameters. 
For the cases studied here, increasing $m$ reduces the pulse-induced deformation of the selected throat-crossing congruence.
Hence it is expected to differ qualitatively from the corresponding memory effects for different wormholes or other compact objects. 
$\mc{B}$-memory thus provides a useful complementary diagnostic of gravitational-wave memory, particularly for extended congruences in curved spacetimes.

 \end{itemize}

Our results indicate that a complete analysis of $\mc{B}$-memory for different classes of congruences (parallel, converging, diverging-- characterised by different initial conditions) should be followed.
Although we have considered static and spherically symmetric wormhole solution here, observational data indicates that most astrophysical systems in our universe undergo rotation. Hence a possible future goal would be to study gravitational memory effect in the case of rotating wormholes \cite{Teo:1998dp, Kar:2024ctd}. 
It would be interesting to see how the memory effect, for example, in a rotating  wormhole would differ from a stationary wormhole as well as black holes. 
Attempt should be made to study GW memory using symmetries at null infinity using the Bondi-Sachs formalism to explore how the Bondi mass varies, as formulated in \cite{Bhattacharya:2022}. 
We shall address these issues elsewhere.
 

\noindent{\bf Data Availability Statement:} No Data associated in the manuscript

\section*{Acknowledgements}
\noindent  Authors would like to thank Sayan Kar and Sumanta Chakraborty for helpful discussions. SB would like to acknowledge IUCAA Pune for providing necessary facilities under the visitor's programme. SG acknowledges grant provided by BIT Mesra (DRIE/SMS/DRIE-05/2023-24/2352).

\section*{Bibliography} 
\bibliography{main}
\bibliographystyle{unsrt}

\end{document}